# Distributed Compilation System for High-Speed Software Build Processes

Geunsik Lim[1], Minho Lee[2], R.J.W.E. Lahaye[3], and Young Ik Eom[4]
CICE[124], Department of Physics[3], Software Center[1]
Sungkyunkwan University[1234], Samsung Electronics[1]
Suwon 440-746, Republic of Korea[1234]
{leemgs[1], zeteman2ya[2], lahaye[3], yieom[4]}@skku.edu, geunsik.lim[1]@samsung.com

*Abstract*— Idle times of personal computers have increased steadily due to the generalization of computer usage and cloud computing. Clustering research aims at utilizing idle computer resources for processing a variable work-load on a large number of computers. The work-load is processed continually despite of the volatile status of the individual computer resources. This paper proposes a distributed compilation system for improving the processing speed of CPU-intensive software compilation. This reduces significantly the compilation time of mass sources by using idle resources. We expect gains of up to 65% against the non-distributed compilation systems.

*Keyword*s— Distributed Computing, Grid Computing, High-Performance Computing, High-Throughput Computing, Super Computing, Distributed Compilation.

## I. INTRODUCTION

Grid computing [1]-[3] can provide services that effectively distributes tasks to suitable resources connected in a network. For example, we can improve the processing speed of intensive scientific calculations by installing work-load management software, which utilizes idle PCs (Personal Computers) of public libraries or administration systems [4]. However, studies into high performance computing [5], [6] still lack research of public computer facilities, which have a lot of idle times.

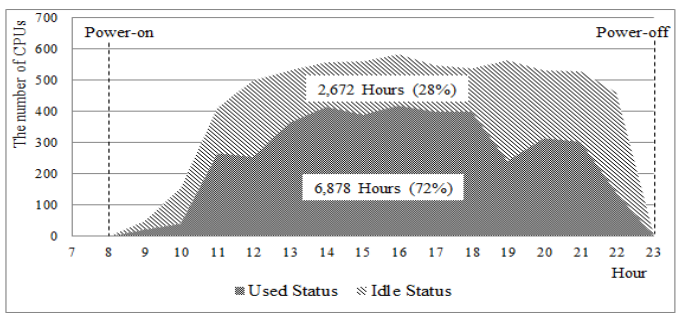

Figure 1. Used time (dark grey) and idle time (light grey) of 300 dual-core PCs on a typical weekday in a university library

The idle times of generic personal computers have increased over time due to generalization of computer usage and cloud computing environments [7] in educational institutions, national administrative agencies, and public institutions. Figure 1 shows an example of used and idle times of 300 dual-core PCs in a university library during a typical weekday. The unused idle times are 28% (2,672 hours) of the times that the PCs are powered on 9,550 hours.

The remainder of this paper is organized as follows. Section II describes the compilation cost due to the evolution of software. Section III addresses the design and implementation of the proposed techniques. Section IV shows the experimental results. Related work is described in Section V. Finally, Section VI concludes the paper.

## II. BUILD COST ANALYSIS OF MOBILE PLATFORM

### A. Software Evolution

The size of recent software is growing due to increased complexity and enhanced functionality of the software. Figure 2 shows the source size growth of popular software from 2009 until 2013. Especially the source size of the Android [8] mobile platform has increased a sevenfold in only 4 years' time.

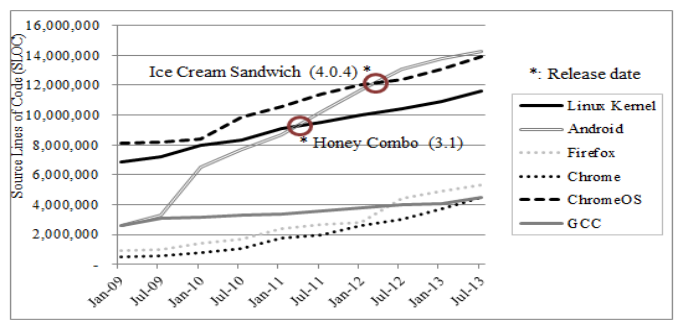

Figure 2. Current state of source size growth of software from January 2009 until July 2013.

### B. Compilation Costs

The tremendous growth of software causes an explosion of the source code lines, which comes along with high software compilation costs. Figure 3 describes the time cost needed to build a large mobile platform such as Android 4.2.2. Compilation costs account for 67 percentages (34 minutes) of the total cost of execution (51 minutes).

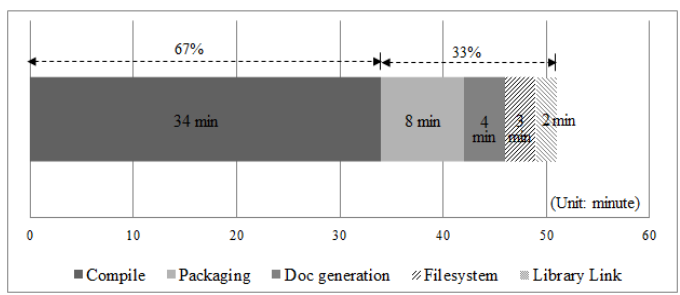

Figure 3. Use time per work for building mobile platform

This research was supported by Basic Science Research Program through the National Research Foundation of Korea (NRF) funded by the Ministry of Education, Science and Technology (2010-0022570).

It is therefore important that we study distributed compilation technology to speed up the compilation of source codes. We can use previous studies [1], [4], [6], [9] to learn how to utilize effectively various hardware devices in a distributed network.

### III. DESIGN AND IMPLEMENTATION OF DISTCOM

The Distributed Compilation System (DistCom) is a technique designed for implementing a distributed compilation platform in order to speed up the compilation of large software by using the resources of idle PCs.

#### A. System Architecture of DistCom

Figure 4 shows the diagram of the entire DistCom system. The proposed DistCom system consists of three components:

- Server and client model, to control distributed PC resources connected via the network.
- DistCom manager, for scheduling distributed PC resources.
- Cross-compiler infrastructure, to support heterogeneous architecture when compiling source codes.

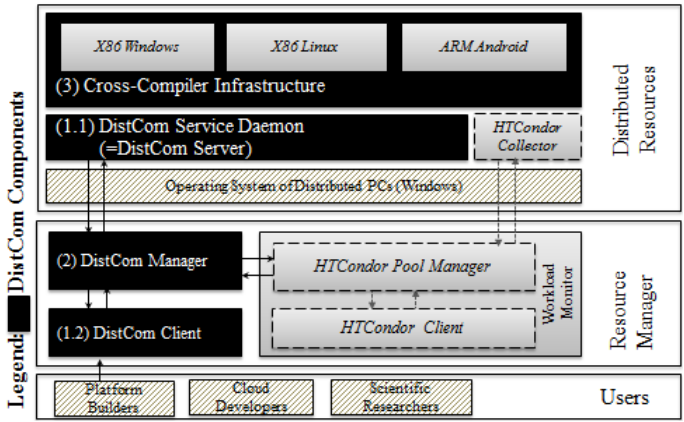

Figure 4. System architecture of the DistCom system

#### B. Distributed Server and Client Model

Figure 5 shows the operation diagram of the server and client model [9], [10] for executing source code as a distributed compilation. When users execute the compilation command via the *(1.2) DistCom Client*, the *(2) DistCom Manager* acquires the information of the distributed PC resources such as the work-load status and the user's access status. Then, the *(2) DistCom Manager* distributes the compilation commands to the distributed PCs to run the distributed compilation of the software code.

The *(2) DistCom Manager* uses a *checkpoint/restart* mechanism [11] to minimize speed degradation, where object files are the atomic level for check pointing. The *checkpoint/restart* is the ability to save the state of a compiling source code so that compilation can later resume on the same or a different distributed PC from the moment at which it was checkpointed. The approach is practical and useful because the proposed system supports retry mechanism that executes recompilation based on the object file units whenever compilation failure of a distributed PC happened during the distributed compilation. The *(2) DistCom Manager* distributes work-load to the *(1.1) DistCom Service Daemons* depending on work-load status of the PCs in the distributed network. Whenever the *(1.1) DistCom Service Daemon* of the distributed PC finishes compilation tasks in its queue, the *(2) DistCom Manager* distributes the new tasks to the *(1.1) DistCom Service Daemon*.

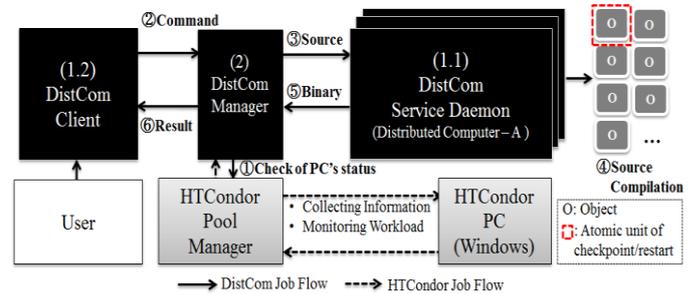

Figure 5. Server and client flowchart of DistCom for the distributed network

#### C. CPU Scheduling of Distributed PC Resources

Our proposed system finds a suitable idle PC resource through the *HTCondor Pool Manager* [12], [13]. When DistCom uses the distributed PC resources, the *(2) DistCom Manager* uses two methods to control the CPU resources in order to optimize the compilation performance of running tasks on the distributed computer.

- *Dedicated resource scheduling*: This operates when the CPU resources are always idle. It is a useful and practical policy in case a user does not use all CPU power of a distributed PC resource. The existing *Distcc* [14] allocates one work-load per PC sequentially regardless of the number of CPUs.
- *Shared resource scheduling*: This shares CPU usage of a distributed PC when its CPU is runnable. This method is useful in case of few idle PCs and the CPU usage of PCs is not high.

The performance of our *dedicated resource scheduling* is similar to the existing study [14]. However, our proposed technique drastically improves the performance by utilizing the *checkpoint/restart* mechanism and by maximizing the CPU usage in multi-core system. When ④*Source Compilation* of Figure 5 is executed, the minimum number of the object files is equal to the number of available CPUs.

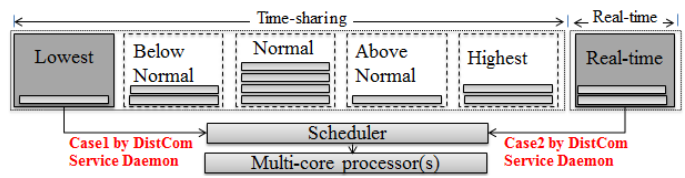

Figure 6. User-aware CPU resource scheduling for improving the performance of the compilantion on idle PCs

Figure 6 shows the CPU sharing method when the *(1.1) DistCom Service Daemon* runs with the *shared resource scheduling* method. To avoid degrading the processing speed during user's work period, the *(1.1) DistCom Service Daemon* runs compilation as a task of *real-time priority* to monopolize a CPU resource in case that the user does not use the PC resource. On the other hand, this run compilation as a task of *lowest priority* to use available CPU usage in case that user accesses the PC resource. Therefore, we do not need any modification of distributed computer system because *(1.1) DistCom Service Daemon* dynamically controls the scheduling priority of itself in the distributed computers. Scheduling priority of *(1.1) DistCom Service Daemon* is dependent on user's input device access such as keyboard, mouse, and touch-screen. The *(1.1) DistCom Service Daemon* decides whether or not it must allocate the next task continually after finishing the allocated compilation jobs. At this time, Criteria of task allocation are the user's input device access, the work-load of remote PC resource, and scheduling cost of task.

Figure 7 shows the three state transition diagram of task to speed up the processing speed by controlling the distributed compilation effectively. First, *Reject* is to deny allocation of task. Second, *Stop* is to break allocation of task to the PC resource because of the user's access. Finally, *Finish* is to complete running tasks normally. *(2) DistCom Manager* manages all jobs with two task queues to separate either dedicated resources or shared resources.

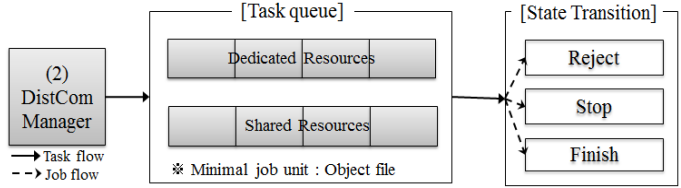

Figure 7. Three state transition flowchart of the task maneaged by DistCom manager for the high-performance distributed compilation

### D. Cross-Compiler Infrastructure to Support Distributed CPU Architectures

Binary files from a compilation must be independent of the CPU architecture of the remote PC resource. In other words, the compiled binary files [15] have to be executed at specified target devices. Figure 8 shows the cross-compiler infrastructure [16] for generating executable binary files for a system other than the one on which the compiler is running. The existing system does not handle different operating systems between the distributed build environment and target environment [10], [17], [18]. Therefore, the *Heterogeneous CPU Mapper* of the proposed cross-compiler infrastructure connects a source code up to the target machine code after probing the operating system structure of the distributed PCs.

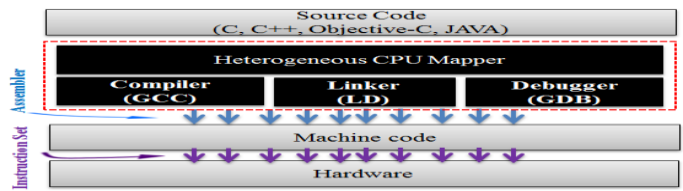

Figure 8. Cross-compiler diagram for handling the distributed compilation environment and the target environment

## IV. EXPERIMENTAL RESULT

We used distributed idle PC resources to evaluate the effectiveness of DistCom in a real environment. When we executed a distributed compilation, we set the maximum available computers to 9 machines (CPU: Intel Core2Duo, MEM: DDR2 1G, Network Interface Card: Intel 100 Mbps Ethernet Controller) and the *dedicated resource scheduling* policy as the default CPU scheduling policy. Figure 9 shows the time costs of the distributed compilation. From our experiment, the time cost to build this mobile platform source is reduced by 65 percent (33 minutes).

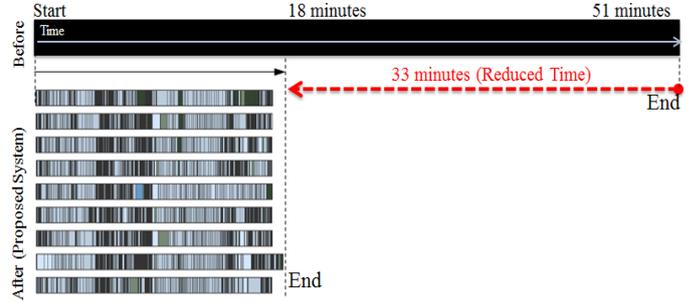

Figure 9. Compilation time by a single computer (top time line) and of a distributed dedicated system with 9 computers (the nine lines below). The compilation time is reduced by 33 minutes from 51 to 18 minutes.

Figure 10 shows the breakdown of compilation cost. We verified that 25% is consumed by the *Network Speed*, 30% by the *Computing Power of PCs*, and 45% by the *CPU Scheduling Method*.

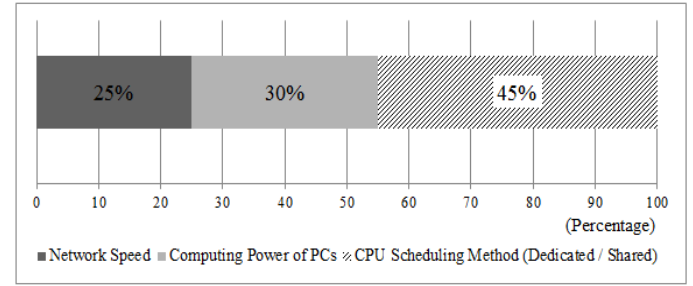

Figure 10. The breakdown of distributed compilation cost of mobile platform source code

In addition, we differently set experimental environment with 5, 10, 15, 20, 25, 30 machines (CPU: Intel Core2Duo, MEM: DDR2 1G, Network Interface Card: Intel 100 Mbps Ethernet Controller) to evaluate performance improvement difference depending on the number of available PC resources. The Intel Ethernet Controller supports a theoretical maximum bandwidth of 100 Mbps.

Figure 11 shows how PC resources connected by network for building mobile platform source affect the processing speed of the distributed compilation. From our experiment, *dedicated resource scheduling* method was faster than *shared resource scheduling* method by 40% on average. From our analysis, we found that compilation processing performance of *shared resource scheduling* method largely depends on the CPU usage of PC resource in comparison with *dedicated resource scheduling* method. Moreover, in case those available PC resources are more than 10 distributed PCs, the compilation speed of *dedicated resource scheduling* method was improved against a high-performance computing server (8-Core Intel Xeon E5 Processor, 12GB memory). Theoretically the performance of 8 machines would be similar to the 8-Core PC. From our analysis, we found that the performance loss of 2 PCs results from network speed and low computing power of the distributed PCs.

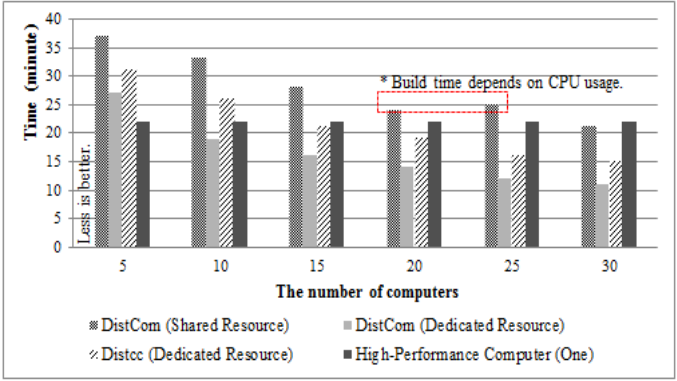

Figure 11. Comparison of the compilation processing performance between the distributed PCs and the high-performance computer server

Figure 11 also shows the comparison of the compilation processing performance between Distcc [14] and DistCom (our system). The existing Distcc [14] allocates one work-load per PC without one work-load per CPU and does not support *shared resource scheduling* like DistCom. However, our proposed system executes multi-core aware distributed compilation that allocates one work-load per CPU after calculating the number of CPUs. From our experiments, we figured out that multi-core aware *dedicated resource scheduling* technique was more effective than the existing Distcc.

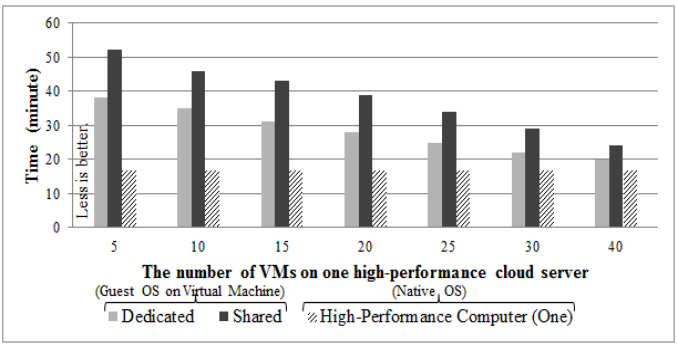

Figure 12. Performance comparison between a cloud server (Guest OSes on KVM virtual machine) and one high-performance server (native operating system)

Figure 12 shows the experimental result when we executed our proposed system on a cloud computing environment. The cloud computing environment consisted of up to 40 Windows XP VMs (Virtual Machines) as Guest OSes after installing KVM virtual machine [19] on a high-performance cloud server (40-Core Intel Xeon E7 Processor, 32GB memory). The results show that the *dedicated resource scheduling* method with 40 VMs consumed 20 minutes and the high-performance cloud server on its native operating system consumed 17 minutes. The 3 minutes difference in performance is caused by the emulation operation [20] of the KVM virtual machine. This demonstrates that in the case of many idle VMs, our proposed system is as effective as one high-performance computer.

Figure 13 shows the results when our proposed system is executed with *Ccache* [21], [22] storing redundant files to random access memory at compilation time. *Ccache* is a program that caches the output of the compilation in order to speed up the second-time compilation, which can significantly speed up the overall recompiling time. When we ran the compilation of a mobile platform source together with *Ccache*, we reduced the compilation time by about 10%. Further analysis showed that the effect of the *Ccache* [23] is correlated with the memory shortage of the distributed PC resources and with the physical memory capacity for caching at compile time.

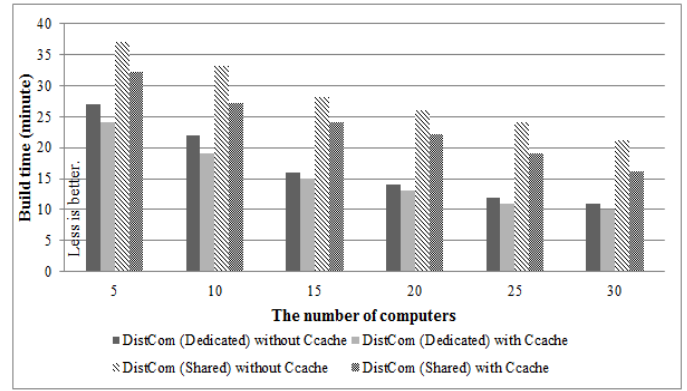

Figure 13. Performance comparison between with and without Ccache

## V. RELATED WORK

In this section, we discuss existing distributed compilation schemes, as well as existing high-performance computing, which served for us as a prelude to the proposed DistCom system.

### A. Existing Works on High-Performance Computing

*Ccache* [21] improves a speed-up of the second-time compilation by storing the output of C/C++ compilation. This technique only focuses on the improvement of recompiling time with available memory. In other words, *Ccache* only handles random access memory without consideration of the available CPU resources. *Distcc* [14] speeds up compilation by distributing compilation tasks across a network of participating hosts. This technique does not handle a *shared resource scheduling* method like our proposed system and is not aware of multi-core CPU environments.

### B. Existing Works on High-Throughput Computing

Berkeley Open Infrastructure for Network Computing (*BOINC*) [24] supports a high performance distributed computing platform for the enormous processing power of personal computers around the world. *BOINC* uses the unused GPU (Graphics Processing Unit) cycles as well as the unused CPU cycles on a computer for scientific computations. However, *BOINC* does not handle distributed compiling tasks. *HTCondor* [25], [26] supports a high-throughput computing framework for coarse-grained distributed parallelization of computationally intensive tasks. However, also *HTCondor* does not handle distributed compilation tasks and is not aware of multi-core CPU environments.

Our proposed system handles high-performance computing as well as high-throughput computing such as a hybrid computing concept. Moreover, our techniques support multi-core aware scheduling without any modifications to the distributed system to manage shared resources as well as dedicated resources.

## VI. Conclusion and Future Work

Idle computer resources connected by a network are more ubiquitous than ever before and it is therefore important to work out a technique of distributed systems, which utilizes unused computer devices. In this paper, we proposed a distributed compilation system, which consists of a distributed server and client model, a resource manager for scheduling distributed computers, and a cross-compiler infrastructure to support heterogeneous architectures. We have verified that our proposed DistCom system can significantly improve compilation speeds using existing idle PC resources by proposing a distributed compiler system of compatible heterogeneous CPU architectures. Moreover, the proposed DistCom system minimizes performance degradation of the distributed compilation by executing resource scheduling of remote computers based on object file units.

In the future, we plan to study a network-aware task scheduling technique considering the physical network speed [27] to distribute tasks and a task migration algorithm [11] to migrate distributed tasks to another idle PC resource.

## Acknowledgment

We thank Seong-Tae Jhang and Hyosuk Kim for their feedback and comments, which helped improve the contents and presentation of this paper.